\documentstyle[referee]{l-aa}
\begin{document}
\thesaurus{02(02.13.2; 02.18.8; 12.12.1)}
\title{Magnetic fields and large scale structure in a hot Universe}
\subtitle{III. The Polyhedric Network}
\author{E. Battaner \inst{1}, E. Florido \inst{1} \and J.M. Garcia-Ruiz \inst{2}}
\institute{Dpto. Fisica Teorica y del Cosmos. Universidad de Granada. Spain \and Instituto Andaluz de Geologia Mediterranea, CSIC-Universidad de Granada. Spain}
\maketitle
\begin{abstract}
We provide a new tool to interpret the large scale structure of the
Universe. As suggested in Paper II, energy density filaments could
have  been
produced by subjacent magnetic flux tubes when the Universe was
dominated by radiation. In more recent time epochs, small scale
filaments have evolved in a complicated way, but large scale filaments
have probably survived and should be identified with present observed
matter filaments of clusters. A primordial magnetic field lattice made
up of magnetic flux tubes would then have originated a matter
lattice made up of cluster filaments. Taking into account some
restrictions in the magnetic field configuration we speculate what
type of matter network this would have generated. The simplest lattice
is made up of octohedra connected at their vertexes, reminding us of a
structure of superimposed egg cartons. The vertexes would correspond
to large superclusters, which would therefore have eight filaments
emerging from them. Some observational evidence is found in the Local
Supercluster, which is spider-like, with eight legs.

\end{abstract}

\section{Introduction}

In Paper I (Battaner, Florido and Jimenez-Vicente 1997) and Paper II
(Florido and Battaner 1997) we  suggested that primordial magnetic
flux tubes have generated radiative energy density filaments throughout the
radiation dominated era. As filaments are very frequently present in
the observed large scale structure today, it is here considered that
contemporary matter filaments could be identified from these early radiative
filaments.

Under this assumption, present large structures would have inherited the
topological structure imposed by the properties of magnetic
fields. This is the goal of this paper, i.e. by assuming that
filaments are magnetically induced condensations of matter and that
they are the pieces with which the Universe is made up, to identify what
observational properties large scale structures would possess.

Filaments are often found in many astrophysical systems, such as the
Sun and the interstellar medium, and are often interpreted as being
magnetically induced. We propose here a similar interpretation for
galaxy cluster filaments.

The model presented in Papers I and II, considers a time period
between Annihilation and Equality; therefore, the matter, radiation
and magnetic field configuration which had been predicted cannot
strictly be compared with present large scale structures. To make such
a comparison valid, the model should be extended to the present,
considering epochs in which the evolution of  filamentary
structures is much more complicated. But the development of a single
model, suitable for so many different epochs is not at present
practical. It is however possible and tempting to qualitatively
predict what kind of structures would now arise from those theoretically
predicted pre-Equality structures. There are two main arguments which
make this prediction reasonable:

-As discussed in paper II, the further evolution of the structures is
subject to three effects which were considered negligible in earlier
epochs: viscosity and heat conduction as the fluid becomes imperfect,
the amplification of magnetic fields by dynamo effects or ejections from galaxies,
and non-linear effects. However, all these processes only affect small
scale structures. Large scale structures, therefore, can be considered
unaltered in recent epochs, and have evolved in a simple way, just
being diluted by expansion, as described, for example, in an Appendix
to Paper I.

-At Equality, filamentary distributions of radiation and matter had
already formed. By Recombination, matter was free to further concentrate
around previous filamentary potential wells. We must therefore
consider the evolution of these early matter structures, even
independently of the subjacent magnetic fields which created them. The
evolution of pre-recombination matter structures is a subject in which
considerable experience has been attained in recent years, and is
considered, for instance in the interpretation of CBR
anisotropies. Again, the evolution of very large scale structures is
much simpler, as it is linear. Magnetic fields could evolve in
different ways and might even affect the evolution of matter distribution,
though probably not in a wholly unpredictable way.

It should not be forgotten, on the other hand, that our analysis in Papers
I and II, was developed for a hot particle fluid, and in particular for
photons before Equality. However the equations could describe actual
structures at present, if hot dark matter were dominant.

We are therefore aware that our discussion is qualitative and
speculative, but nevertheless interesting, as it could open new ways of
interpreting large scale structures. Let us therefore assume that
the present large scale structures are larger than but essentially identical
in shape to the parent pre-Equality filamentary structures.

\section{Magnetic restrictions to the large scale structure}

Independently of the epoch in which primordial magnetic fields were
generated, they must fulfil a restriction: $\nabla \cdot \vec{B}
=0$. Magnetic field lines are either straight lines when viewed at
large scale or they form loops. The first possibility must be excluded
if the Cosmological Principle is maintained. 

Loops can be made from filaments, with plane polygons being the
simplest possibility. Three-dimensional structures made up of polygons
are polyhedra. From the observational point of view, the polyhedric
nature of the Universe is far from demostrated, but this
possibility is by no means in disagreement with observations
(Broadhurst et al., 1990; Einasto et al 1997a,b). We should look for
the simplest polyhedric structures compatible with the absence of
sources and the loss of magnetic field lines.

The Cosmological Principle would require either isolated polygons or
polyhedra with random orientations or period like structures with the
basic polyhedra in contact, forming a network. As isolated structures
are not suggested by observations, we concentrate on the network
structure. Assuming the structure is periodic and three-dimensional, a
crystallographic approach seems reasonable.

Of course, we are not proposing that the Universe is a pure
crystal. More irregular and imperfect forms would actually be
produced, reminding us more of a form structure than a crystal, but a
perfect network is an adequate zero-order theoretical description. A
network is a description of the large scale structure commonly found in
the literature. However, the edges of the network polyhedra now have a
direction, that of the parent magnetic field.

Magnetic field lines connect filaments and there exists in principle the
possibility that they have a complicated contour travelling through
different basic polyhedra. However, the simplest option is to close the loop
within only one polyhedrum or even, within a single face.

The simplest structure would be that when all edges, all faces and all
vertexes are equivalent. This implies that the magnetic flux at any of
the vertexes vanishes. Any magnetic field line reaching a vertex
through one of the edges must quit the vertex through one of the other
equivalent edges. Any arrow of the magnetic field entering a vertex
must exist. This is a severe restriction as it implies that the number
of edges converging at a vertex must be ``even''.

The simplest case is when the basic polyhedra has four edges
converging at a vertex (2 is impossible and 3 is ``forbidden''). For
example, out of the five regular polyhedra, the octahedron has this
property. The icosahedron also has an even number of edges at a
vertex but it is less simple.

Octahedra do not fill up the whole space, i.e. they cannot produce a
Bravais lattice. Some minerals have lattices made up of a combination
of octahedra and tetrahedra, but this possibility is ruled out here
because only three edges converge at a tetrahedron vertex.

When considering octahedra, or any other less simple polyhedra, in
contact, they may share a filament merged from two contacting
edges. In this case, we assume that both edges have the same direction
of magnetic field, because otherwise reconnection of magnetic field
lines would take place. This greatly restricts the possible ``a
priori'' ways of putting the basic polyhedra in contact.

We therefore conclude that the simplest lattice is made up of regular
octahedra. In the next section we will consider which octahedron
lattice is compatible with all the magnetic and simplicity requirements.

There are also some interesting possibilities which do not fulfil all of our
requirements, in particular the requirement that all edges and vortexes should be equivalent. Let us
describe one of the most interesting, even if it was disregarded as
the simplest one.  There are only fourteen kinds of simple space
lattices (Bravais lattices). Among them, those named body centred
structures are the most likely to optimize the above restriction of
four edges converging at a node of the unit cell (see figure 1).

 Note that the same is true for tetragonal and orthorhombic body-centred structures, that is, the configuration of directionality accepts dilation operation along the orthogonal axis of the cell.

\begin{figure}
\caption[]{ Body centred cube, showing convergent and divergent vertices and the central point}
\end{figure}

In this particular body-centred cube all edges are equivalent. Not all
vertexes are equivalent. Convergent vertices are those in which
magnetic fluxes from the three convergent edges converge. Divergent
vertexes are those in which magnetic fluxes from the three convergent
edges diverge. To ensure $\nabla \cdot \vec{B}=0$, the straight line
joining a vertex and a central point, must support a magnetic flux
three times the flux of one of the edges. When these individual body
centred cubes are assembled to make a complete tessellation, it is
easy to calculate that not all vertexes and central points are
equivalent. Taking the magnetic flux in each individual edge as unity,
the magnetic flux entering (or exiting) a vertex is 24. However, at
the central point it  is only 12. Vertexes and central points are not
equivalent, nor are convergent and divergent vertexes equivalent. The lattice containing both types of vertexes is, however, simple, like a salt crystal.

\section{The ``Egg-Carton'' Universe}

Therefore, eliminating the body-centred cubic solution, we have looked
for octahedric structures with all vertexes and edges completely
equivalent. After trials, the only possibility was found to be the one shown in
fig. 2. This consists of a primitive cubic lattice in which an
octahedron is located at each lattice node and then exploded until it
connects with its six nearest neighbours through the vertexes. In this
configuration eight edges converge at  any one vertex. In addition, all the edges and vertexes are equivalent, supporting the same magnetic flux. Note again that this configuration maintains the topological properties when a dilation operation along the ortogonal axis converts the octahedron shape into a bipyramid. This lattice is built up from octahedra joining only at their vertexes.

This lattice reminds us of a structure of superimposed ``egg-cartons'',
with the spaces for the eggs representing the voids constituting the large scale structure.

\begin{figure}
\caption[]{ Lattice of octahedra contacting at their vertexes}
\end{figure}

It is difficult to find observational evidence to check this prediction, due to the relative scarcity of data available today about the large-structure of the Universe. Our edges are super-dense photon filaments which after recombination become matter filaments. At those places where edges or filaments converge we would have still larger concentrations of matter. Filaments are made up of superclusters (e.g. Coma) or simply connect superclusters (Haynes \& Giovanelly 1986, Tago, Einasto \& Saar 1984).

It is an observational fact that filaments are predominant (Gregory \&
Thomson 1978; Joeveer \& Einasto 1978; Tully \& Fisher 1978;
Lapparent, Geller \& Huchra 1986) with voids in between, and most of
them connect superclusters (Gott, Weinberg \& Melott 1987).

 But the prediction here would be that {\it eight} filaments join each other in a supercluster or in a non-filamentary region in which the matter concentration is specially high. Are these superclusters, from which eight  filaments diverge, in the actual structure?

Einasto (1992) and Einasto et al. (1984) compare the structure of the Local Supercluster with a spider. Spiders have eight legs and Einasto's spider is no exception. Examining figure 18 in Einasto's (1992) review it is observed that precisely eight filaments diverge. Einasto's spider has eight legs.

\begin{acknowledgements}We have held valuable discusssions with the crystallographer M. Rodriguez-Gallego. \end{acknowledgements}

\end{document}